\documentclass[acmsmall]{acmart}

\AtBeginDocument{%
  }

\usepackage{multirow}
\usepackage{tablefootnote}
\usepackage{tcolorbox}
\usepackage{colortbl}
\usepackage{mdframed}
\usepackage{xcolor}
\usepackage{soul}

\definecolor{linecolor}{RGB}{105,105,105} 
\definecolor{bgcolor}{RGB}{234,234,234}

\newmdenv[
  linecolor=linecolor,
  leftline=true,
  topline=false,
  bottomline=false,
  rightline=false,
  linewidth=2pt,
  innerleftmargin=10pt,
  innerrightmargin=10pt,
  innertopmargin=5pt,
  innerbottommargin=5pt,
  backgroundcolor=bgcolor
]{leftbar}

\setcopyright{cc}
\setcctype{by}
\acmDOI{10.1145/3728963}
\acmYear{2025}
\acmJournal{PACMSE}
\acmVolume{2}
\acmNumber{ISSTA}
\acmArticle{ISSTA086}
\acmMonth{7}
\received{2024-10-31}
\received[accepted]{2025-03-31}

\begin{document}

\title{Can LLMs Replace Human Evaluators? An Empirical Study of LLM-as-a-Judge in Software Engineering}

\author{Ruiqi Wang}
\orcid{0009-0000-9336-0598}
\affiliation{%
  \institution{Harbin Institute of Technology, Shenzhen}
  \city{Shenzhen}
  \country{China}
}
\email{24s151158@stu.hit.edu.cn}

\author{Jiyu Guo}
\orcid{0009-0001-7055-0767}
\affiliation{%
  \institution{Harbin Institute of Technology, Shenzhen}
  \city{Shenzhen}
  \country{China}
}
\email{220110126@stu.hit.edu.cn}

\author{Cuiyun Gao}
\orcid{0000-0003-4774-2434}
\authornote{Corresponding author.}
\affiliation{%
  \institution{Harbin Institute of Technology, Shenzhen}
  \city{Shenzhen}
  \country{China}
}
\email{gaocuiyun@hit.edu.cn}

\author{Guodong Fan}
\orcid{0000-0003-2031-0615}
\affiliation{%
  \institution{Harbin Institute of Technology, Shenzhen}
  \city{Shenzhen}
  \country{China}
}
\email{guodong.fan@126.com}

\author{Chun Yong Chong}
\orcid{0000-0003-1164-0049}
\affiliation{%
  \institution{Monash University Malaysia}
  \city{Bandar Sunway}
  \country{Malaysia}
}
\email{chunyong@ieee.org}

\author{Xin Xia}
\orcid{0000-0002-6302-3256}
\affiliation{%
  \institution{Zhejiang University}
  \city{Hangzhou}
  \country{China}
}
\email{xin.xia@acm.org}

\renewcommand{\shortauthors}{Ruiqi Wang, Jiyu Guo, Cuiyun Gao, Guodong Fan, Chun Yong Chong, Xin Xia}

\begin{abstract}
Recently, large language models (LLMs) have been deployed to tackle various software engineering (SE) tasks like code generation, significantly advancing the automation of SE tasks. However, assessing the quality of these LLM-generated code and text remains challenging. The commonly used Pass@\(k\) metric necessitates extensive unit tests and configured environments, demands a high labor cost, and is not suitable for evaluating LLM-generated text. Conventional metrics like BLEU, which measure only lexical rather than semantic similarity, have also come under scrutiny. In response, a new trend has emerged to employ LLMs for automated evaluation, known as LLM-as-a-judge. These LLM-as-a-judge methods are claimed to better mimic human assessment than conventional metrics without relying on high-quality reference answers. Nevertheless, their exact human alignment in SE tasks remains unexplored.

In this paper, we empirically explore LLM-as-a-judge methods for evaluating SE tasks, focusing on their alignment with human judgments. We select seven LLM-as-a-judge methods that utilize general-purpose LLMs, alongside two LLMs specifically fine-tuned for evaluation. After generating and manually scoring LLM responses on three recent SE datasets of code translation, code generation, and code summarization, we then prompt these methods to evaluate each response. Finally, we compare the scores generated by these methods with human evaluation. The results indicate that output-based methods reach the highest Pearson correlation of 81.32 and 68.51 with human scores in code translation and generation, achieving near-human evaluation, noticeably outperforming ChrF++, one of the best conventional metrics, at 34.23 and 64.92. Such output-based methods prompt LLMs to output judgments directly, and exhibit more balanced score distributions that resemble human score patterns. Finally, we provide insights and implications, concluding that current state-of-the-art LLM-as-a-judge methods can potentially replace human evaluations in certain SE tasks.
\end{abstract}



\maketitle

\section{Introduction\label{intro}}
Since BERT \cite{DBLP:conf/naacl/DevlinCLT19} and GPT \cite{radford2018improving}, pre-trained language models (PLMs) have been widely used in various natural language processing (NLP) tasks, such as machine translation and text summarization. With the scaling of PLM parameters, the concept of large language models (LLMs) has been proposed. Featuring up to hundreds of billions of parameters, LLMs emerge new capabilities absent on smaller models \cite{DBLP:journals/tmlr/WeiTBRZBYBZMCHVLDF22}, beyond solving simple linguistic tasks. These capabilities include, but are not limited to, instruction following and multi-step reasoning, enabling LLMs to simulate human experts and achieve state-of-the-art performance in certain domains. Software engineering (SE) is one of the specialized domains that benefits from this trend. Many researchers and companies either emphasize their LLMs' strong coding performance \cite{DBLP:journals/corr/abs-2303-08774,DBLP:journals/corr/abs-2403-05530}, or develop specialized code LLMs. For instance, DeepSeek-Coder-V2 \cite{DBLP:journals/corr/abs-2406-11931} correctly generates code for 75.3\% instructions in HumanEval \cite{DBLP:journals/corr/abs-2107-03374} and MBPP \cite{DBLP:journals/corr/abs-2108-07732} with 236B parameters, second only to GPT-4o. Qwen2.5-Coder \cite{hui2024qwen2} achieves 88.4\% Pass@1 on HumanEval with merely 7B parameters. 

However, there has been limited progress in evaluating LLM-generated content for SE. The commonly used Pass@\(k\) metric executes the first \(k\) generated code snippets on human-curated unit tests. While Pass@\(k\) evaluates the code's functional correctness accurately, it has several limitations, such as requiring comprehensive unit tests and manual configuration of test environments. What is more, Pass@\(k\) is unable to evaluate code from non-functional aspects, such as readability and adherence to good practice, nor can it be used to judge text-generating SE tasks like code summarization and code review \cite{DBLP:journals/corr/abs-2308-10620}. Therefore, some SE datasets \cite{DBLP:journals/corr/abs-2409-10280,DBLP:conf/emnlp/YanTLCW23} resort to use conventional metrics such as BLEU \cite{DBLP:conf/acl/PapineniRWZ02} and CodeBLEU \cite{DBLP:journals/corr/abs-2009-10297}, which also have downsides like inability to perform multi-aspect evaluation and requiring human-annotated reference answers. These metrics also focus on lexical rather than semantic similarity, making the evaluation results questionable.

Meanwhile, NLP researchers attempt to apply LLMs to evaluate the quality of LLM-generated content, known as LLM-as-a-judge \cite{DBLP:conf/nips/ZhengC00WZL0LXZ23}. While human effort remains reliable for evaluation and for curating the reference answers in datasets, it is both slow and expensive, defeating the purpose of automatic evaluation. Therefore, researchers prompt or train LLMs to align with human preference, as an attempt to replace human evaluators. 
Since both code and text can be viewed as sequences of tokens, LLM-as-a-judge methods can be potentially adopted on SE tasks. Unfortunately, current meta-evaluation benchmarks feature a limited number of simple coding tasks as they mostly target NLP tasks. The lack of test samples and the insufficient task difficulty create a gap between benchmarking on existing datasets and real-world SE scenarios, where the instructions, code, and responses are usually more complex and varied.

To bridge the gap, we conduct an empirical study to apply a range of LLM-as-a-judge methods on realistic SE datasets. Specifically, we select a task for each of the three input-output type combinations, and a recent representative dataset for each task: CodeTransOcean \cite{DBLP:conf/emnlp/YanTLCW23} for Code Translation (Code-Code), ComplexCodeEval \cite{DBLP:journals/corr/abs-2409-10280} for Code Generation (Text-Code), and CodeXGLUE \cite{DBLP:conf/nips/LuGRHSBCDJTLZSZ21} for Code Summarization (Code-Text). Their corresponding papers only adopt conventional metrics like Exact Match (EM), BLEU, and CodeBLEU. We randomly sample 50 instructions from each dataset, and three out of 12 code LLMs to generate responses for each instruction. For each response, we manually assign a score indicating its quality, resulting in a dataset of 450 samples of (instruction, response, score) triplets in total. Then we perform meta-evaluation of different types of LLM-as-a-judge methods by calculating their score alignment with human scores, to validate whether their judgments match human preference in real-world scenarios.

We design the following three research questions (RQs):

\begin{itemize}
    \item \textbf{RQ1: Which LLM-as-a-judge method aligns with human preference better, and do they outperform conventional metrics?} 

    We aim to assess whether various LLM-as-a-judge methods can replace human evaluators due to high human alignment and superior performance to conventional metrics. We select seven methods across embedding-based, probability-based, and output-based categories, along with two LLMs fine-tuned specifically for NLP evaluation along with their base model, and conventional metrics such as BLEU. We compute the correlations between human scores and scores from these methods to indicate how well they align with human preference on the selected SE tasks.

    \item \textbf{RQ2: What are the characteristics of LLM scores, more specifically their alignments with one another and score distributions?} 

    We aim to characterize the score distributions from LLM-as-a-judge methods with their distributions and correlations. Specifically, we measure the correlations among all methods, to determine if similar methods yield similar results, and to assess whether they actually mimic human evaluators beyond merely measuring lexical similarity. We also analyze the score distribution of each method to investigate their ability to generate varied scores. 

    \item \textbf{RQ3: How do LLMs perform when prompted to make pairwise comparisons instead of individual scoring?} 

    Comparing two responses is also a common choice for LLM-as-a-judge methods, with some studies claiming its superiority over scoring individual responses \cite{DBLP:journals/corr/abs-2310-01432}. We conduct similar experiments to evaluate the performance of these methods when LLMs are instead prompted to select a better response from two, or declare a tie. Since embedding-based and probability-based methods cannot perform this ternary classification without scoring each response first, we focus solely on output-based methods in this RQ. 

\end{itemize}

Through answering the RQs, we conclude that:
\begin{itemize}
    \item The human alignments of studied methods heavily depend on the SE tasks. Among them, output-based methods with large LLMs perform best, achieving near-human performance in code translation and generation.
    \item Similar methods yield similar score distributions, most of which differ from those of conventional metrics. The best human-aligning methods demonstrate more balanced and human-like distributions.
    \item Studied methods fail to deliver accurate and consistent comparison results. Output-based methods with large LLMs still provide the highest accuracy, but often yield inconsistent results after swapping the positions of two responses.
\end{itemize}

Our contributions can be summarized as follows:
\begin{itemize}
    \item Our work serves as the first empirical study to investigate applying LLM-as-a-judge methods specifically to SE tasks, with much more difficult code-specific instructions and responses compared to previous studies.
    \item We manually curate a meta-evaluation dataset based on three existing SE datasets for different tasks, to evaluate human alignment of LLM-as-a-judge methods.
    \item We explore how different LLM-as-a-judge methods prefer to score responses, and discuss the findings and possible implications for their future studies and applications in SE.
\end{itemize}

The rest of the paper is organized as follows: Section \ref{literature} introduces research relevant to code LLMs, SE task evaluation, and notable LLM-as-a-judge methods. Section \ref{overview} offers more details in different categories of LLM-as-a-judge methods. Section \ref{experiment} presents the overall study design. Section \ref{result} records the experimental results and analyzes our findings. Section \ref{discussion} analyzes score explanations as a case study and possible future directions based on our findings. Section \ref{conclusion} concludes the paper.
\section{Related Work\label{literature}}
\subsection{Code LLMs for SE}
LLMs are large-scale PLMs. Some of them are instruction-tuned to follow instructions in human language. In this paper, we do not distinguish between PLMs and LLMs, and use LLMs to refer to pre-trained Transformers \cite{DBLP:conf/nips/VaswaniSPUJGKP17} in general.

While many general-purpose LLMs demonstrate satisfying performance on SE tasks, especially code generation, there are many LLMs pre-trained specifically for code-related tasks. CodeBERT \cite{DBLP:conf/emnlp/FengGTDFGS0LJZ20} is one of the earliest attempts to pre-train a Transformer on both code and text data. It is an encoder-only model with 125M parameters, pre-trained on over 8M datapoints from CodeSearchNet \cite{DBLP:journals/corr/abs-1909-09436}. CodeT5 \cite{DBLP:conf/emnlp/0034WJH21} is an encoder-decoder Transformer with up to 770M parameters, pre-trained with a denoising sequence-to-sequence objective on the same dataset. 
UniXcoder \cite{DBLP:conf/acl/GuoLDW0022} supports encoder-only, decoder-only, and encoder-decoder modes, allowing abstract syntax trees (ASTs) as input after transforming ASTs into sequences.

Recently, larger decoder-only LLMs have been increasingly popular in generation tasks. Codex \cite{DBLP:journals/corr/abs-2107-03374} is a series of GPT-based LLMs with up to 12B parameters, achieving a Pass@1 score of 28.81\% on HumanEval. CodeLlama \cite{DBLP:journals/corr/abs-2308-12950} is another LLM family with up to 70B parameters from Meta AI, trained from Llama 2 \cite{DBLP:journals/corr/abs-2307-09288} to follow human instructions. 
DeepSeek-Coder \cite{DBLP:journals/corr/abs-2401-14196} is a family of LLMs with up to 33B parameters, supporting both normal generation and fill-in-the-middle (FIM). Its successor, DeepSeek-Coder-V2 \cite{DBLP:journals/corr/abs-2406-11931}, is a mixture-of-experts (MoE) LLM with 16B or 236B parameters, claiming to have GPT-4 \cite{DBLP:journals/corr/abs-2303-08774} level performance at a Pass@1 score of 90.2\% on HumanEval.

\subsection{SE Benchmarks and Metrics}
Many SE benchmarks focus solely on code generation, where LLMs generate code for the given requirements and function signatures. HumanEval \cite{DBLP:journals/corr/abs-2107-03374} is one of the most adopted code generation benchmarks, featuring 164 human-curated Python problems. It uses Pass@\(k\) as the evaluation metric. 
MBPP \cite{DBLP:journals/corr/abs-2108-07732} is another popular benchmark with 974 Python problems, aiming at entry-level development. APPS \cite{DBLP:conf/nips/HendrycksBKMAGB21} is a much larger Python benchmark with 10000 problems, ranging from being solvable in one-line to presenting substantial challenges in algorithms. 
ClassEval \cite{DBLP:journals/corr/abs-2308-01861} challenges LLMs with 100 class-level code generation problems in Python, and measure class-level and method-level Pass@\(k\). 

Some benchmarks target other SE tasks. CodeReviewer \cite{DBLP:conf/sigsoft/LiLGDJJMGSFS22} aims at three tasks in the code review process: commit quality estimation, reviewer comment generation, and code editing. CodeXGLUE \cite{DBLP:conf/nips/LuGRHSBCDJTLZSZ21} supports 10 SE tasks such as code summarization and code search. ComplexCodeEval \cite{DBLP:journals/corr/abs-2409-10280} collects code from influential GitHub repositories for 4 tasks such as code generation and unit test generation. These benchmarks all evaluate responses with conventional metrics including Exact Match, Edit Similarity, BLEU, and CodeBLEU instead of Pass@\(k\), even for code-generating tasks. CRUXEval \cite{DBLP:conf/icml/GuRLSS024} evaluates LLMs from other aspects such as code understanding and execution with 800 short Python functions for input or output predictions. It requires LLMs to output assert statements to obtain Pass@\(k\) scores. 

However, limited efforts are made to curate meta-evaluation benchmarks to test evaluation metrics, as most datasets only contain instructions and reference answers, without responses of different quality or human-annotated scores. NoFunEval \cite{DBLP:journals/corr/abs-2401-15963} designs six evaluation aspects, including functional correctness and non-functional aspects like latency and maintainability. It tests whether LLMs can improve code based on a specific aspect or select the better of two code snippets from that perspective. CodeUltraFeedback \cite{DBLP:journals/corr/abs-2403-09032} evaluates LLMs' alignment with human evaluation from five non-functional code aspects like instruction following and coding style.

\subsection{LLM-as-a-Judge in NLP}
\subsubsection{Embedding-Based Methods}
Some researchers obtain contextual token representations of the response and reference answer using encoder-only LLMs, and compute pairwise similarity to obtain the score. BERTScore \cite{DBLP:conf/iclr/ZhangKWWA20} calculates Recall, Precision and F\(_1\) score based on token representations obtained from BERT \cite{DBLP:conf/naacl/DevlinCLT19}. It also applies inverse document frequencies (IDFs) to reduce the weight of overly common and thus less essential tokens. MoverScore \cite{DBLP:conf/emnlp/ZhaoPLGME19} constructs a transportation cost matrix based on token representations and computes Word Mover's Distance \cite{DBLP:conf/icml/KusnerSKW15}. CodeBERTScore \cite{DBLP:conf/emnlp/Zhou0AN23} is a code-specific adaptation of BERTScore with CodeBERT, approximating functional correctness and human scores with F\(_3\) and F\(_1\) scores respectively. While these methods match contextual embeddings instead of n-grams, unlike many conventional metrics, they still measure how a response resembles the reference answer.

\subsubsection{Probability-Based Methods}
Since more LLMs come with decoders, it becomes possible to use generating probabilities for evaluation. BARTScore \cite{DBLP:conf/nips/YuanNL21} assumes that BART \cite{DBLP:conf/acl/LewisLGGMLSZ20} is more likely to generate a higher-quality response. It uses the probability of BART generating a given response as the score. GPTScore \cite{DBLP:conf/naacl/FuNJ024} applies a similar approach with 19 LLMs of sizes from 80M to 175B, supporting both reference-free and reference-based evaluation from multiple aspects. FFLM \cite{DBLP:conf/emnlp/0003RLZ23} is a reference-free method designed to evaluate the faithfulness of summaries. It calculates the probabilities of generating the summary with and without the original text as posterior and prior probabilities respectively. FFLM assumes that a faithful summary has higher posterior than prior probability, and calculates their difference as the score.

\subsubsection{Output-Based Methods} 
While the above methods usually align with human evaluation better than conventional metrics, they do not explain their scores or support certain closed-source LLMs that do not provide probabilities or representations. Output-based methods prompt LLMs to output the judgments, and do not require access to their internal implementations. G-Eval \cite{DBLP:conf/emnlp/LiuIXWXZ23} utilizes Chain-of-Thought (CoT) \cite{DBLP:conf/nips/Wei0SBIXCLZ22} to request evaluation steps, samples multiple scores and then averages them as the final score. ChatEval \cite{DBLP:conf/iclr/ChanCSYXZF024} assigns different personas to several LLM agents, asking them to discuss and select a better response from two. 

Some researchers construct training sets to fine-tune LLMs instead of designing prompting or inference strategies. InstructScore \cite{DBLP:conf/emnlp/XuWPSFWL23} is fine-tuned on GPT-4-synthesized data to generate error reports of text from various domains. PandaLM \cite{DBLP:conf/iclr/WangYYZYW0J000024} is fine-tuned on pairwise comparison results and reference answers generated by GPT-3.5, aiming at addressing subjective aspects including conciseness and clarity. X-Eval \cite{DBLP:conf/naacl/LiuSX0CKGH24} has an extra training stage to learn the connections between fine-grained evaluation aspects, allowing evaluating from aspects not seen during training.

However, these methods have not been tested on a sufficient number of challenging SE samples, leaving it unclear whether they achieve reliable human alignment for SE applications.
\section{\label{overview}LLM-as-a-Judge Framework Overview}
\begin{figure}
    \centering
    \includegraphics[width=\textwidth]{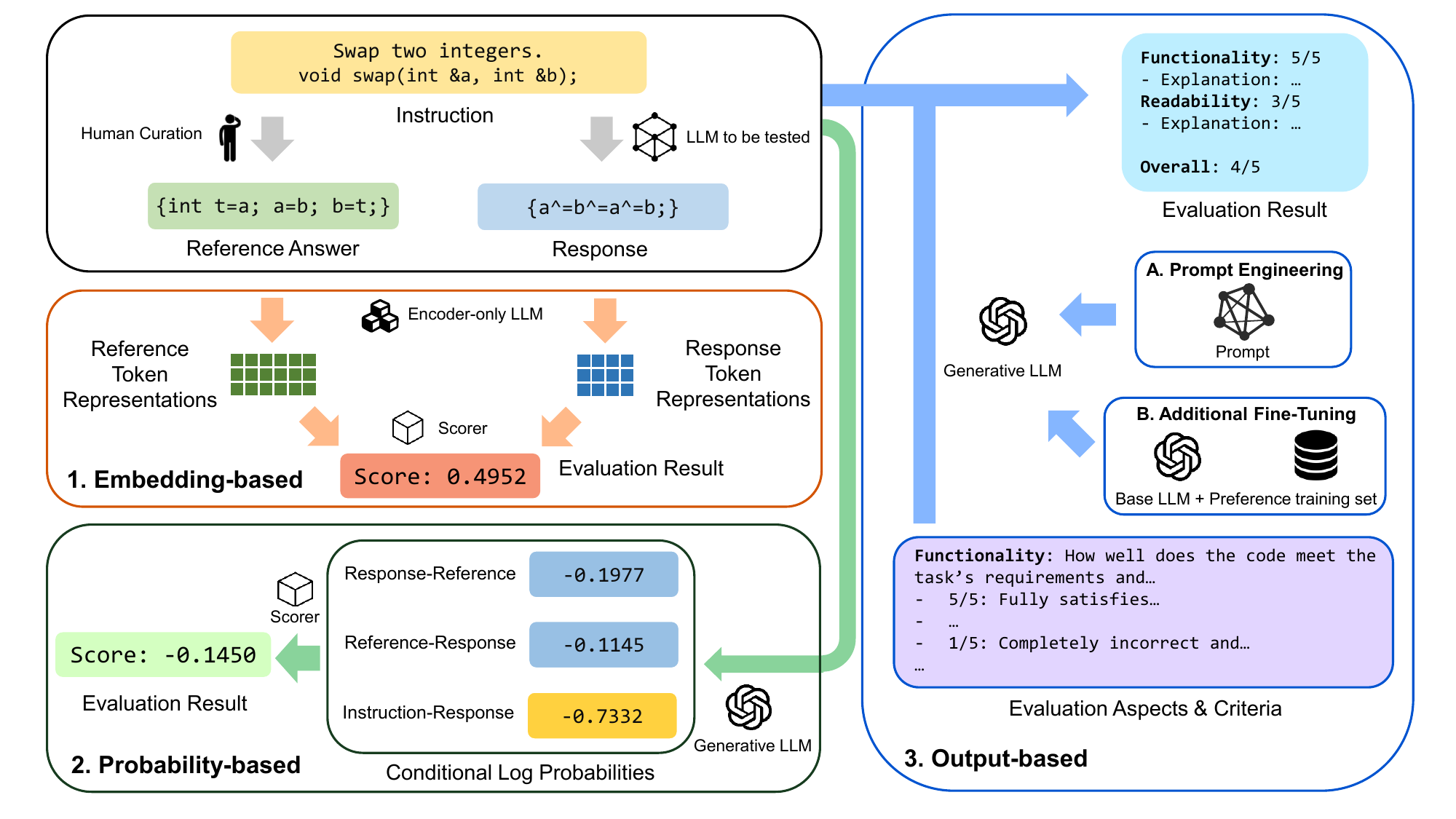}
    \vspace{-2.0em}
    \caption{Overview of different LLM-as-a-judge methods.}
    \label{fig:approach}
\vspace{-1.0em}
\end{figure}

In this section, we offer an overview of existing LLM-as-a-judge methods. As seen in Fig. \ref{fig:approach}, we categorize these methods based on the types of LLM features used\footnote{Our categorization is inspired by \cite{DBLP:journals/corr/abs-2402-01383}. We merge similar categories based on LLM feature types.}, including embedding-based, probability-based, and output-based methods. We denote the instruction (source) as \(src=s_1...s_{|src|}\), the response (target) as \(tgt=t_1...t_{|tgt|}\), and the reference answer as \(ref=r_1...r_{|ref|}\).
\begin{itemize}
    \item \textbf{Embedding-based}: These methods first obtain token representations of the response and reference answer \(f(tgt)=f(t_1)...f(t_{|tgt|})\) and \(f(ref)=f(r_1)...f(r_{|ref|})\) from the LLM encoder \(f\). We then evaluate \(tgt\) 
    via fusing token-wise cosine similarities \(s_{ij}=\frac{f(t_i)\cdot f(r_j)}{\lVert f(t_i)\rVert\lVert f(r_j)\rVert}\). 

    \item \textbf{Probability-based}: The LLM receives an input-output pair \((in, out)\), and returns the conditional log-probability of generating \(out\), i.e. \(\log p(out|in)=\frac 1{|out|}\sum_{k=1}^{|out|}\log p(out_k|in,out_{<k})\). Typical \((in, out)\) combinations include \((src,tgt)\), \((ref,tgt)\), \((tgt,ref)\), and \((none,tgt)\), where \(none\) means no input is provided. We then score \(tgt\) with these log-probabilities. Additional content may be present in the prompt, such as evaluation aspects like clarity.

    \item \textbf{Output-based}: These methods first craft a prompt \(prompt\) with \(src\) and \(tgt\). Depending on the design, \(prompt\) may also feature \(ref\), evaluation aspects and criteria, and evaluation steps. After obtaining the judgment \(jud=\text{LLM}(prompt)\), we extract the final score from \(jud\). Many prompting and inference strategies can also be applied, such as multi-agent and repeated sampling, where multiple scores are often combined using methods like a majority vote or averaging.

    \quad LLMs can also be fine-tuned as specialized judges, usually applied with a single inference pass and no additional strategies. State-of-the-art LLMs like GPT-4 are often used to generate reference judgments for training. In this paper, we discuss the performance of these LLMs, instead of focusing on the detailed training process.

    \quad Unlike embedding-based and probability-based methods, which usually have scoring ranges of \([0, 1]\) and \((-\infty, 0)\) (or \((-\infty, \infty)\)) respectively without rescaling, most output-based methods require LLMs to score on a scale of 1 to 5 or 1 to 10. They can also compare two responses and decide the better one or declare a tie. In our study, we investigate individual scoring in RQ1 and RQ2, and pairwise comparison in RQ3.
\end{itemize}

\section{Study Design\label{experiment}}
In this section, we elaborate on the details of our study design. In our study, we focus on leveraging different types of LLM-as-a-judge methods to evaluate the responses of three SE tasks. We collect instructions and generate responses from representative datasets, then perform human and LLM evaluation on these responses, and analyze their correlations.

\subsection{Datasets and Preprocessing}
\subsubsection{Instruction Collection}
To ensure the difficulty of instructions and to approximate real-world development scenarios, we collect instructions from a recent dataset for each of the three widely-studied SE tasks for our empirical evaluation:

\begin{itemize}
    \item \textbf{Code Translation} is a code-to-code task demanding translating code between two languages while preserving the functionality. It challenges LLMs' skills to understand syntax and library usages in both languages, and to choose replacements when certain functionalities are unavailable in the target language.

    \item \textbf{Code Summarization} is a code-to-text task involving generating a concise and fluent description of a given code snippet. It challenges LLMs' abilities to abstract the code, leaving only critical information about core functionality rather than explaining step-by-step.

    \item \textbf{Code Generation} is a text-to-code task requiring generating a function based on a natural language description and the signature. It tests LLMs' capabilities to breakdown the functional requirement into steps, and to utilize provided dependencies.
\end{itemize}

We select the \textbf{MultilingualTrans}\footnote{Collected from Rosetta Code, \url{https://rosettacode.org/wiki/Rosetta_Code}.} subset of \textbf{CodeTransOcean} \cite{DBLP:conf/emnlp/YanTLCW23} for code translation. CodeTransOcean contains three translation subsets for different purposes, with the MultilingualTrans subset covering eight popular languages with 7545 samples. Compared to previous benchmarks, CodeTransOcean offers more pairs of programming languages of longer code, with the average length of test sets reaching 491 tokens in MultilingualTrans, as opposed to 58 tokens in CodeTrans featured in CodeXGLUE, which only supports Java and C\#. CodeTransOcean evaluates translations with conventional metrics such as Exact Match, BLEU, and CodeBLEU, rather than execution-based metrics like Pass@\(k\), since they require constructing unit tests and testing environments.

We select the code-text subset of \textbf{CodeXGLUE} \cite{DBLP:conf/nips/LuGRHSBCDJTLZSZ21} for code summarization, which is a filtered version of CodeSearchNet\footnote{Collected from public GitHub repositories.} \cite{DBLP:journals/corr/abs-1909-09436}. Initially developed for code search, i.e. retrieving relevant code based on natural language queries, CodeSearchNet contains two million code snippets in six programming languages accompanied by docstrings. These docstrings come from the associated function documentation and serve as summaries. CodeXGLUE removes samples with syntactically incorrect code, or docstrings that are either non-English, overly lengthy, or too short. After filtering, 14918 Python samples and 10955 Java samples remain, along with samples in four other programming languages. CodeXGLUE uses BLEU to evaluate generated summaries.

We select \textbf{ComplexCodeEval}\footnote{Collected from GitHub repositories.} \cite{DBLP:journals/corr/abs-2409-10280} for code generation. ComplexCodeEval is a benchmark with 3897 and 7184 Java and Python samples respectively, supporting four tasks: code generation, code completion, unit test generation, and API recommendation. Compared to previous benchmarks, ComplexCodeEval provides comprehensive supplemental material for each code snippet, including functional dependencies, timestamps, and unit tests. It expects LLMs to learn project-specific dependencies beyond standard library or popular third-party APIs. ComplexCodeEval evaluates code-generating tasks with conventional metrics such as Edit Similarity, BLEU, and CodeBLEU. 

When training, validation, and test sets are available, we only adopt the test set for our evaluation. To ensure the accuracy of manual evaluation, we limit the programming languages to Java, Python, C, and C++ according to the human evaluators' expertise. Following previous work's \cite{DBLP:journals/corr/abs-2310-03304,DBLP:journals/corr/abs-2405-01535} context length of 4096 tokens, we also apply length limits of 1536, 1536, and 1024 tokens\footnote{Measured with OpenAI's Tiktoken, \url{https://github.com/openai/tiktoken}, with GPT-4o's vocabulary o200k\_base.} for instructions, responses\footnote{Here we limit the length of reference answers instead of actual responses generated in the next step.}, and output-based judgments respectively, removing samples with lengthy instructions or reference answers. We sample 50 instructions from each filtered dataset, resulting in 150 instructions in total.

\subsubsection{Response Generation}
We deploy 12 recent code LLMs with different sizes from seven families shown in Table \ref{tab:llm_list} from Hugging Face \cite{DBLP:journals/corr/abs-1910-03771}. We generate responses using these LLMs with vLLM \cite{DBLP:conf/sosp/KwonLZ0ZY0ZS23} on an Ubuntu 20.04 server with two Intel Xeon Platinum 8276L CPUs, four NVIDIA A100-40GB GPUs, and 256 GB RAM. For each instruction, we randomly select three LLMs to respond, yielding three responses \(A,B,C\). For pairwise comparisons, we create three response pairs \((A,B), (A,C),(B,C)\), and another three pairs \((B,A),(C,A),(C,B)\) in order to check if studied methods yield consistent judgment after reversing the order within a response pair. Thus, we obtain 150 responses and 300 response pairs per task, resulting in 450 responses and 900 pairs in total. 

\begin{table}[]
    \centering
    \caption{Selected LLMs for response generation, sorted by their release date.}
    \vspace{-1.0em}
    \begin{tabular}{c|c|c|c}
    \toprule
    LLM Family & Developer & Size & Date \\ 
    \midrule
    CodeLlama-Instruct \cite{DBLP:journals/corr/abs-2308-12950} & Meta AI & 7/13/34B & 2023.8 \\ 
    DeepSeek-Coder \cite{DBLP:journals/corr/abs-2401-14196} & DeepSeek AI & 1.3/6.7/33B & 2023.11 \\ 
    MagiCoder-S-DS \cite{DBLP:conf/icml/0003W0D024} & UIUC \& THU & 6.7B & 2023.12 \\ 
    Codestral-v0.1\tablefootnote{Announced at \url{https://mistral.ai/news/codestral/}.} & Mistral AI & 22B & 2024.5 \\ 
    DeepSeek-Coder-V2-Lite \cite{DBLP:journals/corr/abs-2406-11931} & DeepSeek AI & 16B & 2024.6 \\ 
    CodeGeeX4-ALL \cite{DBLP:conf/kdd/ZhengXZDWXSW0LS23} & Zhipu AI \& THU & 9.4B & 2024.7 \\ 
    Qwen2.5-Coder \cite{hui2024qwen2} & Alibaba & 1.5/7B & 2024.9 \\
    \bottomrule
    \end{tabular}
    \label{tab:llm_list}
\vspace{-0.5em}
\end{table}

\begin{table}[]
    \centering
    \caption{Contextual information provided for each task in response generation.}
    \vspace{-1.0em}
    \begin{tabular}{c|c}
        \toprule
        Task & Contextual Information \\ 
        \midrule
        Code Translation & Original Code \\
        Code Summarization & Original Code \\
        Code Generation & Signature, Description, Dependencies \\
        \bottomrule
    \end{tabular}
    \label{tab:context}
\vspace{-1.0em}
\end{table}

As part of the prompt, contextual information in Table \ref{tab:context} is provided for LLMs. The full prompts are available in our repository \cite{replication-package}. LLMs are permitted to generate at most 3072 tokens, two times the maximum reference answer length, to minimize the need for truncation.

After preliminary experiments, we discover that many reference summaries in CodeXGLUE are in fact incorrect. Therefore, we require the reference summary to at least have 15 tokens, reselect the instructions, and manually examine each instruction. We also find that in code generation, selected LLMs struggle to generate interpretable code because they cannot use dependencies effectively, as the only available dependency information in ComplexCodeEval is their names, which makes human evaluation almost impossible to yield meaningful scores. Consequently, we reselect the instructions with at most five dependencies to reduce difficulty, and augment the dependency information with GPT-4o\footnote{We use the 2024-08-06 version for all experiments. The prompt for augmentation is available in our repository.}, prompting it to extract the signature from the reference answer\footnote{The authors of ComplexCodeEval extract dependency names from reference answers as well, and we follow their practice to utilize reference answers.} and generate a short description for each dependency. We manually examine the descriptions to ensure that no other information about the reference answers is included. Responses are generated with the updated dependency information as well as other contextual information.

\subsubsection{Manual Evaluation}
For manual evaluation, we design two evaluation aspects per task to guide human evaluators, enabling more fine-grained assessment without overwhelming evaluators with too many aspects and complicated criteria. The first aspect assesses the response's alignment with the instruction, e.g. Consistency with Code for summarization, requiring the summary to capture the code's core functionality. The second aspect judges the response's intrinsic quality, e.g. Readability \& Idiomatic Usage for translation, demanding the responded code to be both readable and follow common coding styles in the target language. We also curate the criteria for each integer score ranging from 1 to 5 for both aspects. In general, a 5-point response is near perfect, a 4- or 3-point response contains minor or major issues but still makes sense, and a 2- or 1-point response is practically useless. Below, we show an example of Readability aspect and its corresponding criteria for code summarization as an example, while the remaining aspects and criteria can be found in our repository:

\begin{tcolorbox}
\textbf{Readability:} How clear, concise, and fluent is the summary in describing the code’s function?

\textbf{- 5/5:} Extremely clear, concise, and well-structured; very easy to understand.

\textbf{- 4/5:} Mostly clear and concise, with minor readability issues.

\textbf{- 3/5:} Understandable but may contain some unclear or awkward phrasing.

\textbf{- 2/5:} Hard to follow due to unclear language or poor structure.

\textbf{- 1/5:} Very confusing, with significant language or structural issues.
\end{tcolorbox}

Two human evaluators with expertise in the chosen programming languages are involved in judging each of the 450 responses. During manual evaluation, we provide the corresponding instruction and the reference answer along with the response to be evaluated. Each evaluator is required to score both aspects before assigning an overall score, which is not necessarily the average of the former. The final human score for each response is the average of overall scores from two evaluators\footnote{The two human evaluators reach a high level of agreement, achieving Spearman's \(\rho\) of (83.07, 75.42, 74.20), Pearson's \(R\) of (85.86, 79.70, 73.74), and Kendall's \(\tau\) of (72.26, 63.40, 62.57) on code translation, generation, and summarization respectively.}. For pairwise comparison, we calculate the absolute difference between the final human scores of two responses in a pair, declaring a tie when the difference is smaller than 0.5\footnote{The value is chosen so that ties occur for about a third of the response pairs for each task.}, or deciding the higher-scored response is better otherwise. 

\subsection{Selected Methods}
\subsubsection{Conventional Metrics}
We choose five popular conventional metrics, each requiring the response \(tgt\) and the reference answer \(ref\) but not the instruction. We verify if these metrics align better or worse with human evaluation compared to LLM-as-a-judge methods.
For details about Recall, Precision, and F\(_n\) scores, please refer to their original papers.

\textbf{BLEU} \cite{DBLP:conf/acl/PapineniRWZ02} calculates modified n-gram precision (\(n=1,2,3,4\)) for \(tgt\) and \(ref\), and applies a brevity penalty to penalize overly short responses.

\textbf{ROUGE-L} \cite{lin2004rouge} measures the length of the longest common subsequence \(\text{LCS}\) between \(tgt\) and \(ref\). It computes the F\(_1\) score based on LCS.

\textbf{METEOR} \cite{DBLP:conf/acl/BanerjeeL05} matches tokens in \(tgt\) and \(ref\), and computes the F\(_3\) score based on the number of matched tokens. It also penalizes fragmented alignment by counting the number of contiguous match chunks in \(tgt\).

\textbf{ChrF++} \cite{DBLP:conf/wmt/Popovic17} computes F\(_2\) scores using character n-grams (up to 6-grams) and token n-grams (up to 2-grams). The average character F\(_2\) score and the average token F\(_2\) score are then averaged to produce the final score.

\textbf{CrystalBLEU} \cite{DBLP:conf/kbse/EghbaliP22} is specifically designed to measure code similarity. It removes the most common n-grams in a corpus from \(tgt\) and \(ref\), as these trivial n-grams can obscure meaningful differences between them, before calculating BLEU score. For each task, we use the test set from its corresponding dataset as the corpus, including all instructions and reference answers.

We implement the first four methods with Hugging Face Evaluate, and the last with the CrystalBLEU package. For methods with replaceable tokenizers, we substitute them with OpenAI Tiktoken with o200k\_base vocabulary because the built-in tokenizers are usually not designed for code. 

\subsubsection{Embedding-Based Methods}
We choose two methods based on embedding, i.e. token representations of the response \(tgt\) and the reference answer \(ref\). We use UniXcoder \cite{DBLP:conf/acl/GuoLDW0022} in place of BERT or other non-code LLMs as our encoder, due to its ability to process both code and text.

\textbf{BERTScore} \cite{DBLP:conf/iclr/ZhangKWWA20} calculates pairwise token similarity between \(tgt\) and \(ref\) with token representations, and obtains the average Recall and Precision, which are combined into the F\(_1\) score as the final score. BERTScore also applies inverse document frequencies (IDFs) as token weights.

\textbf{MoverScore} \cite{DBLP:conf/emnlp/ZhaoPLGME19} proposes to use Word Mover's Distance \cite{DBLP:conf/icml/KusnerSKW15}, measuring semantic dissimilarity as the minimum cost flow between n-gram representations, which is the IDF-weighted average of token representations. For each \(n\), it constructs a cost matrix for each n-gram in \(tgt\), and flow requirements based on IDF. The final score is the minimum cost to establish such a flow.

\subsubsection{Probability-Based Methods}
We select two probability-based methods. These methods may take at least two of the following as input: instruction \(src\), response \(tgt\), and reference \(ref\), plus supplementary information like evaluation aspects\footnote{Aspects are identical to those in human evaluation.} \(a\). We use davinci-002 here, since later OpenAI models only return probabilities of newly generated tokens instead of prompt tokens.

\textbf{GPTScore} \cite{DBLP:conf/naacl/FuNJ024} simply uses the sequence log probability \(\log p(tgt|src,a)\) as the score according to their paper, which is the average of all token log probabilities. However, their code instead uses the harmonic mean of \(\log p(tgt|ref,a)\) and \(\log p(ref|tgt,a)\). To mitigate this difference, we additionally include \(src\) in both conditions, i.e. using \(\log p(tgt|ref,src,a)\) and \(\log p(ref|tgt,src,a)\).

\textbf{FFLM} \cite{DBLP:conf/emnlp/0003RLZ23} is a reference-free metric that obtains both the prior probability \(P(tgt)\) and the posterior probability \(P(tgt|src)\). It claims that high-loss (low-probability) tokens contribute more to low-quality content, thus assigning a higher weight to them. FFLM also introduces the prefix probability \(P(tgt|tgt:src)\) by prepending \(tgt\) to \(src\), assuming that the prefix increases the generating probability if \(tgt\) is inconsistent with \(src\). These three probabilities are fused into the final score.

\subsubsection{Output-Based Methods}
We select two methods: G-Eval and BatchEval, which apply different inference strategies, in addition to a control group (Vanilla) with no strategies applied, to assess if these strategies improve alignment with human evaluation for general-purpose LLMs. Unless otherwise stated, we use GPT-4o for these methods.

We also include a supervised fine-tuning (SFT) group, with two LLMs fine-tuned for NLP evaluation, along with their base LLMs without fine-tuning, to determine if fine-tuning for NLP evaluation also enhances human alignment in SE evaluation. 

We provide only the instruction \(src\), response \(tgt\), and evaluation aspects\footnote{Aspects are identical to those in human evaluation.} in the prompt. For the detailed prompts, please refer to our repository. Note that these methods can also perform pairwise comparison, where we include both responses in the prompt. 

\textbf{Vanilla} performs inference once with greedy decoding (temperature set to 0), where LLMs score each aspect first before assigning the final score. For pairwise comparison, LLMs compare on each aspect and then make the final decision. We use DeepSeek-Coder-V2-Lite locally, and DeepSeek-V2.5 and GPT-4o via API.

\textbf{G-Eval} \cite{DBLP:conf/emnlp/LiuIXWXZ23} requires the LLM to generate the evaluation steps first and embeds it into the prompt, followed by 20 inference passes with a high temperature of 1.0 and averaging the scores. Following their practice, we prompt the LLM to return the score first with a limit of 20 generated tokens. Judgments without a score are discarded. For pairwise comparison, we consider comparison results as scores of 1 or -1 when one response is better, or 0 for a tie. If the absolute value of the average score is less than 0.7\footnote{The value is chosen so that ties occur for about a third of the response pairs.}, we declare a draw, otherwise considering one response better.

\textbf{BatchEval} \cite{DBLP:conf/acl/YuanFLWPWH024a} performs multi-round scoring. In each round, it first batches all responses and then scores each batch in one inference pass. During batching, it diversifies the scores of responses in each batch, so that the LLM can learn an unbiased score distribution for more accurate scoring. We follow their practice by setting temperature to 0.2, batch size to 10, and number of rounds to 5.

\textbf{SFT} involves two LLMs fine-tuned for NLP evaluation, Auto-J \cite{DBLP:conf/iclr/LiSYF0024} and Prometheus-v2-BGB-8x7B \cite{DBLP:journals/corr/abs-2405-01535}, as well as their base LLMs Llama2-13B-Chat \cite{DBLP:journals/corr/abs-2307-09288} and Mixtral-8x7B-Instruct \cite{DBLP:journals/corr/abs-2401-04088}. We apply the default prompt template of each judge LLM to itself and its base LLM, and exclude the evaluation aspects for Auto-J and Llama-2-13B-Chat since Auto-J's template does not provide a place for aspects. We perform greedy decoding locally with temperature 0.

We only consider the final verdict (score or comparison result) in our meta-evaluation and discard the explanations. For verdict extraction, we use G-Eval's code for itself, and, for all other methods in this category, we set several rules to match with regular expressions, such as "Overall: X" and "[[X]]" where X is the non-negative final score, or comparison result "First", "Second", or "Draw". If no valid verdict is found or the extracted score exceeds 10, which we consider invalid, we assign a score of -1 or a comparison result as draw as a penalty.

\subsection{Meta-Evaluation}
Meta-evaluation refers to the process of evaluating different evaluation metrics. For the default method of individual scoring, we meta-evaluate the metrics via their correlation with human scores, including Spearman's \(\rho\), Pearson correlation coefficient \(R\), and Kendall's \(\tau\). For pairwise comparison in RQ3, we compute the Accuracy of LLM-generated labels, in addition to the Agreement which checks if an LLM makes the same judgment when two responses in the prompt swap their positions.

For the ease of reading, all correlation coefficients, Accuracies, and Agreements in this paper are multiplied by 100. We also check if the \(p\)-value of each correlation coefficient in RQ1 is smaller than 0.05 to ensure a 95\% confidence interval.

\section{Study Results\label{result}}
In this section, we present experimental results and our analysis to answer the research questions.

\subsection{RQ1: Alignment with Human Scores}
\begin{table}[]
    \centering
    \caption{Experimental results for individual scoring. DS2.5 means DeepSeek-V2.5 while DSC2-Lite means DeepSeek-Coder-V2-Lite. The best alignment in each column is marked bold. The best conventional metric alignment and better results in other categories are underlined. Coefficients with \(p>0.05\) are marked red.}
    \vspace{-1.0em}
    \begin{tabular}{c|ccc|ccc|ccc}
         \toprule
         \multirow{2}{*}{Method} & \multicolumn{3}{c|}{Translation} & \multicolumn{3}{c|}{Generation} & \multicolumn{3}{c}{Summarization} \\

         & \(\rho\) & \(R\) & \(\tau\) & \(\rho\) & \(R\) & \(\tau\) & \(\rho\) & \(R\) & \(\tau\) \\ \midrule

         \rowcolor{gray!40} \multicolumn{10}{c}{\textbf{Conventional Metrics}} \\ \midrule
BLEU        & 31.12 & 28.08 & 22.43 & 58.08 & 55.83 & 41.90 & 19.80 &  24.77 &  16.78 \\
ROUGE-L       & 28.55 & 28.57 & 20.29 & 55.72 & 57.62 & 40.81 & \underline{\textbf{48.45}} &  \underline{\textbf{47.01}} &  \underline{\textbf{35.47}} \\
METEOR      & 22.48 & 31.79 & 15.98 & \underline{\textbf{67.11}} & \underline{65.55} & \underline{49.66} & 38.83 &  40.01 &  28.27 \\
ChrF++        & \underline{31.30} & \underline{34.23} & \underline{22.65} & 64.02 & 64.92 & 46.60 & 47.26 &  44.65 &  33.86 \\
CrystalBLEU & 23.63 & 25.26 & 17.43 & 59.02 & 56.65 & 42.88 & 23.19 &  24.96 &  17.24 \\ \midrule

         \rowcolor{gray!40}\multicolumn{10}{c}{\textbf{Embedding-based}} \\ \midrule
BERTScore   & 27.72 & 32.49 & 19.54 & 41.39 & 44.74 & 30.36 & 21.71 &  21.89 &  15.57 \\
MoverScore  & 28.29 & 26.22 & 19.99 & 46.64 & 47.35 & 33.66 & 31.86 &  29.44 &  22.82 \\ \midrule

         \rowcolor{gray!40}\multicolumn{10}{c}{\textbf{Probability-based}} \\ \midrule
GPTScore    & \underline{33.53} & \underline{34.77} & \underline{25.12} & 46.65 & 45.42 & 35.00 &\textcolor{red}{-13.34} & \textcolor{red}{-15.04} &  \textcolor{red}{-9.28} \\
FFLM        & \underline{34.03} & 29.37 & \underline{25.50} & 29.31 & 29.62 & 21.65 & \textcolor{red}{-2.29} &  \textcolor{red}{-8.71} &  \textcolor{red}{-1.94} \\ \midrule

         \rowcolor{gray!40}\multicolumn{10}{c}{\textbf{Output-based: Vanilla}} \\ \midrule
DSC2-Lite   & \underline{33.10} & \underline{46.26} & \underline{26.56} & \textcolor{red}{15.71} & 28.28 & \textcolor{red}{12.50} &-17.76 & -17.47 & -15.25 \\
DS2.5      & \underline{62.43} & \underline{70.27} & \underline{49.48} & 66.39 & \underline{\textbf{68.51}} & \underline{\textbf{54.74}} &  17.73 & 18.10 & 14.14 \\
GPT-4o       & \underline{70.67} & \underline{79.11} & \underline{57.85} & 54.70 & 57.02 & 43.56 &  24.52 & 23.15 & 19.27 \\ \midrule

         \rowcolor{gray!40}\multicolumn{10}{c}{\textbf{Output-based: Inference strategies}} \\ \midrule
G-Eval       & \underline{68.96} & \underline{77.14} & \underline{52.90} & 60.71 & 63.05 & 46.36 & 23.34 &  26.19 &  17.18 \\
BatchEval   & \underline{\textbf{73.67}} & \underline{\textbf{81.32}} & \underline{\textbf{59.80}} & 59.54 & 63.04 & 48.62 & 22.56 &  22.46 &  18.39 \\ \midrule

         \rowcolor{gray!40}\multicolumn{10}{c}{\textbf{Output-based: SFT}} \\ \midrule
Llama2      &  \textcolor{red}{2.61} &  \textcolor{red}{1.03} &  \textcolor{red}{1.92} & 23.91 & 22.89 & 18.94 &\textcolor{red}{-15.61} & \textcolor{red}{-15.81} & \textcolor{red}{-12.82} \\
Auto-J       & 20.99 & \textcolor{red}{14.43} & 17.45 & 36.53 & 38.92 & 29.79 & \textcolor{red}{-5.13} &  \textcolor{red}{-4.92} &  \textcolor{red}{-4.36} \\
Mixtral     & 24.67 & 34.07 & 19.52 & \textcolor{red}{14.41} & 25.32 & \textcolor{red}{11.18} & \textcolor{red}{-3.97} &  \textcolor{red}{-8.98} &  \textcolor{red}{-3.39} \\
Prometheus  & \underline{32.42} & \underline{39.25} & \underline{26.60} & 29.03 & 40.33 & 23.09 &-17.12 & -17.14 & -14.24 \\ \bottomrule
         
    \end{tabular}
    \label{tab:RQ1_res}
    \vspace{-1.5em}
\end{table}

We use LLM-as-a-judge methods to score individual responses and evaluate their correlation with human scores. Table \ref{tab:RQ1_res} presents the alignment between human scores and scores generated by various methods, including both LLM-as-a-judge methods and conventional metrics.
We notice that the three types of correlation coefficients display similar trends, and make the following discoveries:

\textbf{Current LLM-as-a-judge methods lack generalizability, as they demonstrate drastically different performance in different tasks and scenarios.} In Code Translation, BatchEval reaches the highest human alignment, offering near-human performance at \(\rho=73.67\), \(R=81.32\), and \(\tau=59.80\), while G-Eval, DeepSeek-V2.5, and GPT-4o also reach a high correlation of \(R>70\) or \(\rho>60\), greatly outperforming conventional metrics capped at \(R=34.23,\rho=31.30\). We attribute this to the characteristic of responses and reference answers: LLMs often copy statements from the original code with subtle language-specific modifications as the response. Meanwhile, although the reference answer maintain unchanged core functionality, its exact implementation and behavior might noticeably differ. This presents a disadvantage for reference-based methods including most non-output-based methods and conventional metrics. Output-based methods, however, are designed to work without reference and can utilize LLMs' knowledge of programming languages in evaluation.

On the contrary, LLM-as-a-judge methods struggle to outperform conventional metrics in evaluating code generation outputs and are completely surpassed in evaluating code summarization. For code generation, conventional metrics can reach a mid-high correlation of \(\rho=67.11\), \(R=65.55\), and \(\tau=49.66\), while DeepSeek-V2.5 is the only LLM outperforming them at \(\rho=66.39\), \(R=68.51\), and \(\tau=54.74\) without any additional inference strategies. This can be attributed to the characteristics of the ComplexCodeEval dataset, which emphasizes the usage of complicated dependencies by filling out the correct arguments and calling them at the right time instead of designing sophisticated algorithms. Therefore, a response-reference comparison at the lexical level can offer an insight of the response's quality, while the LLMs' limited understanding of the dependencies fail to provide benefits in evaluation. With that said, for code generation, LLM-as-a-judge methods with large LLMs like GPT-4o are still applicable, since they display similar performance as conventional metrics but provide the benefits of not requiring reference answers. For code summarization, LLM-as-a-judge techniques are completely defeated by conventional metrics, hardly reaching a score of 30 in any correlation coefficient or even demonstrating a negative correlation with human evaluation. Nonetheless, conventional metrics also fail to deliver satisfying alignment with human evaluation, with \(\rho,R<50\) and \(\tau<40\). This is potentially due to the fact that many LLMs try to explain the code step-by-step instead of summarizing the core functionality, which is difficult for these LLM-as-a-judge methods to detect. While conventional metrics can assign low scores to these responses, they have trouble handling paraphrasing, which is common in summaries. It is an interesting future direction to explore new metrics that align with humans for code summarization.

\vspace*{6pt}
\begin{leftbar}
\textbf{Finding 1:} Current LLM-as-a-judge methods demonstrate low generalizability in aligning with human evaluation, outperforming conventional metrics in code translation, performing on par with them in code generation, while being outperformed in code summarization.
\end{leftbar}
\vspace*{6pt}

\textbf{Inference using large LLMs yields the best human alignment across all tasks, while inference strategies only provide marginal improvement.} Embedding-based and probability-based methods underperform output-based methods in most scenarios, capped at \(R=34.77,47.35,29.44\) versus the top performance of the latter at \(R=81.32,68.51,26.19\), and the top performance of conventional metrics at \(R=34.23,65.55,47.01\) in code translation, code generation, and code summarization respectively. Furthermore, embedding-based and probability-based methods require access to internal states, while the API services of many state-of-the-art LLMs only allow access to the final output. Therefore, these methods cannot be applied with such LLMs, limiting their applicability. Based on the low human alignment and limited applicable LLMs, we conclude that embedding-based and probability-based methods are impractical for evaluating SE tasks.

Among the output-based methods, we find that DeepSeek-V2.5 and GPT-4o outperform other LLMs without further training. Although Auto-J and Prometheus 2, trained to match human preference, provide better performance than their base model, with a 5.18\% to 16.03\% increase in Pearson's \(R\), achieving \(R=38.92\) and \(R=40.33\) in evaluating code generation respectively, the overall performance is still inferior. This is likely due to the limited number of parameters, as Auto-J and Prometheus 2 only have 13B and 47B parameters. Another possible reason is the misalignment between evaluating NLP tasks during training, and evaluating SE tasks during inference. Though many NLP training datasets contain programming tasks, they may only present common tasks like code generation and fail to present sufficiently challenging instructions. Unfortunately, to the best of our knowledge, no multi-task human preference training sets for SE task evaluation have been curated so far. Hence, we are unable to investigate LLMs fine-tuned on such SE-specific datasets.

Similarly, current inference strategies, when employed to GPT-4o, produce an inadequate performance boost of \(\Delta R=2.21,6.03,3.04\) at maximum. Despite recent work claiming the effectiveness of scaling inference \cite{DBLP:journals/corr/abs-2408-03314}, we found that existing inference strategies for SE evaluation only bring marginal improvement in human alignment. Moreover, they have different downsides: G-Eval forces LLMs to generate the overall score first, restricting the efficacy of the Chain-of-Thought procedure, while greatly increasing inference cost if the full explanations are needed; BatchEval increases the token count, leading to more expensive inference due to multi-round evaluation. Therefore, greedy decoding remains a viable LLM-as-a-judge solution with satisfactory performance and lower requirements of token count, when equipped with colossal state-of-the-art LLMs.

\vspace*{6pt}
\begin{leftbar}
\textbf{Finding 2:} Among the LLM-as-a-judge methods studied, output-based methods with large state-of-the-art LLMs perform best, regardless of inference strategies.
\end{leftbar}
\vspace*{6pt}

\subsection{RQ2: Score Characteristics}
\begin{table}[]
    \centering
    \caption{Category-wise correlation. \(\rho_\text{conv}\), \(\rho_\text{other}\), and \(\rho_\text{inner}\) are the maximum Spearman's \(\rho\) between the specified category with either conventional metrics, metrics from the other three categories, and other metric(s) from the same category. Coefficients above 50 are underlined, while those above 75 are marked bold.}
    \vspace{-1.0em}
    \begin{tabular}{c|ccc|ccc|ccc}
          \toprule
         \multirow{2}{*}{Category} & \multicolumn{3}{c|}{Translation} & \multicolumn{3}{c|}{Generation} & \multicolumn{3}{c}{Summarization} \\

         & \(\rho_\text{conv}\) & \(\rho_\text{other}\) & \(\rho_\text{inner}\) & \(\rho_\text{conv}\) & \(\rho_\text{other}\) & \(\rho_\text{inner}\) & \(\rho_\text{conv}\) & \(\rho_\text{other}\) & \(\rho_\text{inner}\) \\ \midrule

Embedding-based & \underline{\textbf{81.45}} & 37.81 & \underline{74.83} & \underline{\textbf{79.78}} & 46.74 & \underline{\textbf{84.20}} & \underline{57.07} & 23.32 & \underline{66.61} \\
Probability-based & 32.18 & 28.60 & 31.30 & \underline{63.69} & 46.74 & \underline{60.92} & 32.78 & 23.32 & \underline{63.13} \\
Output-based w/o SFT & 30.60 & 35.64 & \underline{\textbf{90.64}} & 47.25 & 39.56 & \underline{\textbf{88.25}} & 20.04 & 48.21 & \underline{79.43} \\ 
Output-based w/ SFT &  37.14 &  37.81 & 24.25 &  30.96 &  37.52 & 27.44 &  11.21 &  48.21 & 23.44 \\ \bottomrule

    \end{tabular}
    \label{tab:RQ2_res}
    \vspace{-0.5em}
\end{table}

\begin{table}[]
    \centering
    \caption{Correlation between output-based LLM-as-a-judge methods grouped by the sizes of the LLMs they use. Methods are categorized based on the underlying model size: "Small" (using <50B LLMs), which includes all methods from the SFT group and DeepSeek-Coder-V2-Lite from the Vanilla group, and "Large" (using >100B LLMs), including DeepSeek-V2.5 and GPT-4o, the latter used by G-Eval and BatchEval.}
    \vspace{-1.0em}
    \begin{tabular}{c|cc|cc|cc}
         \toprule
         \multirow{2}{*}{LLM Sizes Compared} & \multicolumn{2}{c|}{Translation} & \multicolumn{2}{c|}{Generation} & \multicolumn{2}{c}{Summarization} \\

         & \(\rho_{\min}\) & \(\rho_{\max}\) & \(\rho_{\min}\) & \(\rho_{\max}\) & \(\rho_{\min}\) & \(\rho_{\max}\) \\ \midrule

         Small-Small & -4.10 & 24.25 & -2.74 & 27.44 &  -5.55 & 23.44 \\
         Large-Large & 83.04 & 90.64 & 68.63 & 88.25 &  31.50 & 79.43 \\
         Small-Large &  4.16 & 48.92 & 11.70 & 37.84 & -16.15 & 48.21 \\ \bottomrule
    \end{tabular}
    \label{tab:RQ2_llm}
    \vspace{-1.0em}
\end{table}

\begin{figure}
    \centering
    \includegraphics[width=\textwidth]{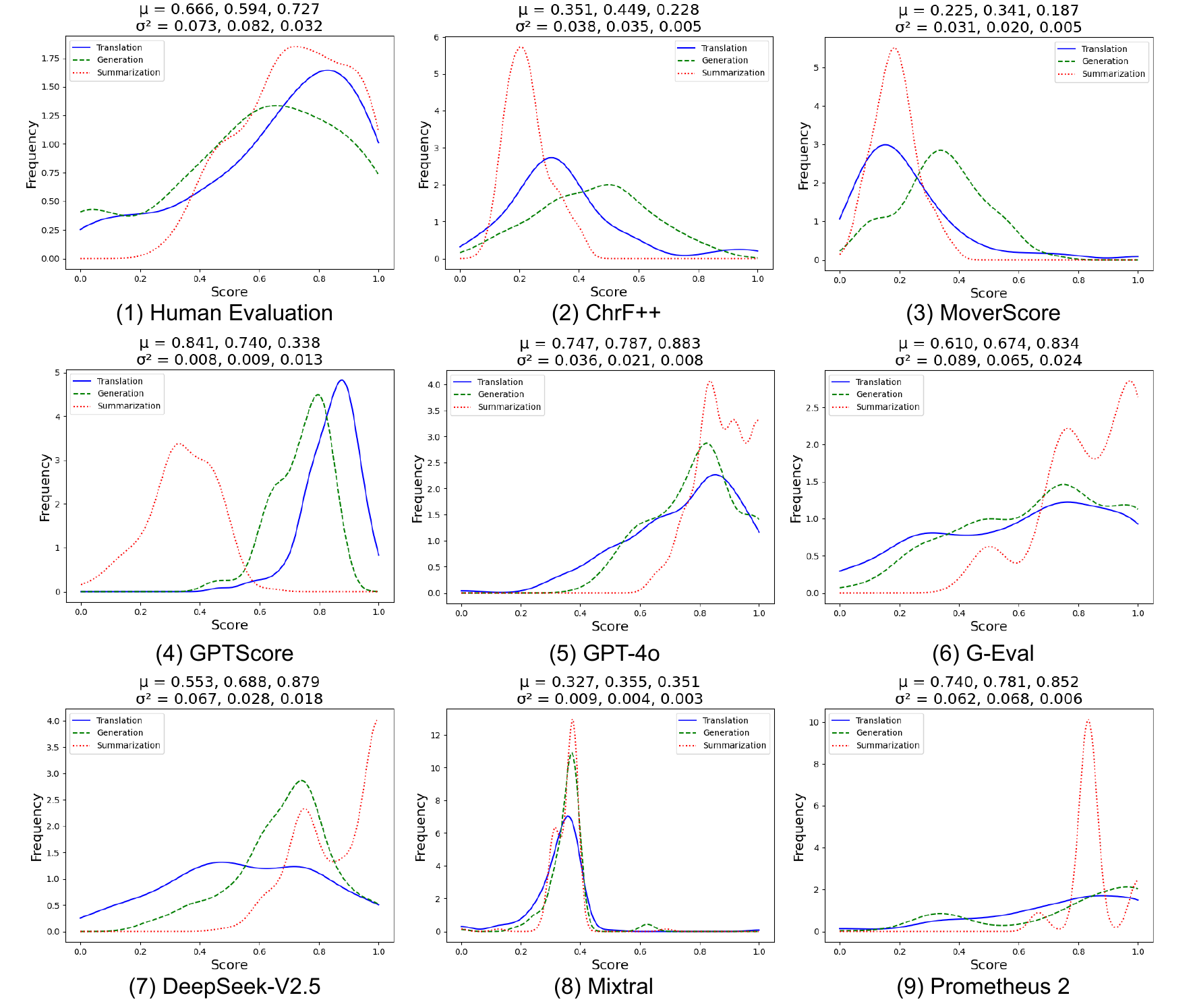}
    \vspace{-2.0em}
    \caption{Score distributions of selected metrics. \(\mu,\sigma^2\) refer to the means and variances of scores for code translation, code generation, and code summarization. All scores are rescaled into range \([0, 1]\).}
    \label{fig:plot}
    \vspace{-1.0em}
\end{figure}

We investigate the score characteristics of various LLM-as-a-judge methods. Table \ref{tab:RQ2_res} shows the maximum correlation between metrics from the same or different categories of LLM-as-a-judge methods, while Fig. \ref{fig:plot} displays the score distributions\footnote{Frequency estimated using Kernel Density Estimation (KDE). All scores rescaled into range \([0,1]\).} of different methods: (1) for manual evaluation, (2) for conventional metrics, (3) for embedding-based methods, (4) for probability-based methods, (5)(6)(7) for output-based methods without SFT, and (8)(9) for output-based methods with SFT. For each distribution, rather than focusing on the specific shape of the curve, we examine whether it is unimodal and note the peak frequency and the corresponding score. 

We make the following discoveries:

\textbf{Most non-SFT LLM-as-a-judge methods have low correlations with those from other categories and high correlations with those from the same category.} In Table \ref{tab:RQ2_res}, we observe that \(\rho_\text{other}<50\) for all categories, meaning that each category demonstrates a unique distribution of scores instead of resembling others. Conversely, \(\rho_\text{inner}>60\) under most non-SFT circumstances, exhibiting a medium to high level of agreement among similar methods. This phenomenon suggests that the mechanics governing each category may significantly influence their score distributions. In contrast, scores from SFT methods correlate poorly even within the same category, likely due to variations in their base LLMs and fine-tuning datasets. Given the high level of disagreement among current fine-tuned LLMs, we argue that selecting an appropriate fine-tuned LLM is crucial for evaluating under specific SE contexts. Otherwise, it may produce entirely unexpected scores. 

\textbf{Output-based methods using large LLMs tend to align well with each other, whereas those using smaller LLMs exhibit low correlations with other methods.} Since output-based methods offer the best human alignment, we further investigate whether LLM size influences correlations between methods by grouping these LLM-as-a-judge methods into those using large LLMs (>100B) and those using small LLMs (<50B). In Table \ref{tab:RQ2_llm}, we observe that methods employing large LLMs achieve high correlations of \(\rho>80\) for code translation and \(\rho>65\) for code generation with each other. These methods use DeepSeek-V2.5 and GPT-4o, and maintain strong alignment despite the difference in LLMs and inference strategies. In contrast, methods using small LLMs yield \(\rho<50\) when compared to methods in the "large" group, and \(\rho<30\) within the "small" group. This pattern reflects a performance gap, as the "large" group align substantially better with human evaluations than the "small" group.

\vspace*{6pt}
\begin{leftbar}
\textbf{Finding 3:} As anticipated, methods within the same category generally exhibit high correlations with each other and low correlations with those in different categories. Among output-based methods, those using large LLMs not only align well with human scores but also show strong correlations with each other.
\end{leftbar}
\vspace*{6pt}

\textbf{Only embedding-based methods resemble conventional metrics.} We discover in Table \ref{tab:RQ2_res} that \(\rho_\text{conv}=81.45\), 79.78, 57.07 for the 3 tasks with embedding-based methods, indicating a high correlation with conventional metrics. As shown in Fig. \ref{fig:plot}, MoverScore from this category exhibits a distribution similar to that of ChrF++, one of the most human-aligning conventional metrics, as both tend to assign low to medium scores to responses. This similarity is anticipated, given that both metrics are designed to assess the similarity between the response and the reference. While MoverScore leverages contextual token representations beyond simple lexical matching, the underlying principles remain fundamentally aligned. On the other hand, distributions from other categories differ markedly from those of ChrF++ as shown by their low \(\rho_\text{conv}\) values, further underscoring their limited resemblance to conventional metrics.

\textbf{The best human-aligning methods closely replicate the distribution of human scores.} As shown in Fig. \ref{fig:plot}, GPT-4o, G-Eval, and DeepSeek-V2.5 in plots (5)(6)(7) demonstrate the highest alignment with human judgment in plot (1), with similar peak frequencies and corresponding scores between 0.6 and 0.8. GPTScore in plot (4) also shows similar peaks for code translation and code generation, though its scores are less evenly distributed compared to these top methods, as seen in the high peak frequency and reduced variance values \(\sigma^2\). In contrast, the underperforming methods, such as ChrF++ in plot (2) and MoverScore in plot (3), peak at much lower scores, resulting in average scores notably below those of human evaluators. Interestingly, Prometheus 2 in plot (9) shows a comparable peak location and a relatively balanced distribution after fine-tuning on Mixtral outside code summarization, yet this does not correspond to a high human alignment.

\vspace*{6pt}
\begin{leftbar}
\textbf{Finding 4:} Only embedding-based methods align closely with conventional metrics, while the most human-aligned output-based methods display more balanced score distributions that mirror human scoring patterns.
\end{leftbar}
\vspace*{6pt}

\subsection{RQ3: Pairwise Comparison versus Individual Scoring}
Since embedding-based and probability-based methods can only score individual responses, we only analyze output-based methods for pairwise comparison. Table \ref{tab:RQ3_res} presents the results.

In general, \textbf{current LLM-as-a-judge methods fail to deliver satisfactory and consistent comparison performance on SE tasks.} For code translation, G-Eval and BatchEval reach the highest Accuracy of 64.67 and 65.33, followed by GPT-4o and DeepSeek-V2.5, and all other methods fall below 50 Accuracy. On code generation, even the best-performing methods struggle to achieve 50 Accuracy, while all methods become completely unusable on code summarization. 

We also evaluate their consistency by reversing the order of the two responses in the prompt to check if methods yield the same comparison results, measured as Agreement. Table \ref{tab:RQ3_res} shows that methods with the highest accuracy on code translation and generation yield extremely low Agreement below 25, indicating poor consistency. Meanwhile, the most consistent methods from the SFT category barely outperform random guessing, where each outcome (selecting a better response or declaring a tie) has an equal \(\frac 13\) chance.

Although unreliable and inconsistent, \textbf{their comparison Accuracy displays a similar trend as in individual scoring.} For the first two tasks, methods applying inference strategies on large LLMs, such as G-Eval and BatchEval with GPT-4o, exhibit the highest Accuracy, followed by DeepSeek-V2.5 and GPT-4o with greedy decoding and no further strategies, though the performance impact of inference strategies is noticeably larger than in individual scoring. For example, BatchEval provides up to +8 Accuracy boost here for GPT-4o compared to +3 in Spearman's \(\rho\) in RQ1 on code translation. Besides, DeepSeek-Coder-V2-Lite, a code LLM with merely 16B parameters, also defeats LLMs fine-tuned to evaluate NLP tasks, but again lags behind large LLMs.

\vspace*{6pt}
\begin{leftbar}
\textbf{Finding 5:} Current LLM-as-a-judge methods exhibit disappointing Accuracy in pairwise comparisons and often yield inconsistent results when the order of two responses is reversed. As with individual scoring, output-based methods using large LLMs achieve the highest Accuracy, yet inference strategies provide a larger performance boost than in individual scoring. However, these strategies do not fully resolve the inconsistency issue.
\end{leftbar}
\vspace*{6pt}

\begin{table}[]
    \centering
    \caption{Experimental results for pairwise comparison, where Acc. means Accuracy and Agr. means agreement. The best result in each column is marked bold. Results higher than 50 are underlined.}
    \begin{tabular}{c|cc|cc|cc}
         \toprule 
         \multirow{2}{*}{Method} & \multicolumn{2}{c|}{Translation} & \multicolumn{2}{c|}{Generation} & \multicolumn{2}{c}{Summarization} \\

         & Acc. & Agr. & Acc. & Agr. & Acc. & Agr. \\ \midrule

         Random guess & 33.33 & 33.33 & 33.33 & 33.33 & 33.33 & 33.33 \\ \midrule

         \rowcolor{gray!40} \multicolumn{7}{c}{\textbf{Vanilla}} \\ \midrule
DSC2-Lite  & 44.67 & 36.00 & 38.67 & 36.67 & 30.33 & 32.00 \\
DS2.5     & \underline{51.00} & 10.67 & 48.33 & 16.67 & 26.67 & 16.67 \\
GPT-4o      & \underline{57.33} & 13.33 & 49.33 & 13.33 & 25.00 & 16.00 \\ \midrule

         \rowcolor{gray!40} \multicolumn{7}{c}{\textbf{Inference strategies}} \\ \midrule
G-Eval      & \underline{64.67} & 17.33 & \underline{\textbf{54.67}} & 15.33 & 34.33 & 32.67 \\
BatchEval  & \underline{\textbf{65.33}} & 21.33 & \underline{52.67} & 24.00 & 36.33 & 38.00 \\ \midrule

        \rowcolor{gray!40}  \multicolumn{7}{c}{\textbf{SFT}} \\ \midrule
Llama2     & 36.00 & \underline{\textbf{78.67}} & 34.67 & \underline{\textbf{72.67}} & 31.67 & \underline{\textbf{56.00}} \\
Auto-J      & 33.33 & \underline{52.00} & 38.33 & 28.67 & 23.33 & 16.00 \\
Mixtral    & 29.00 & 48.67 & 32.00 & 40.00 & 32.67 & 48.67 \\
Prometheus & 33.67 & 35.33 & 37.67 & 31.33 & \textbf{42.00} & 26.00 \\ \bottomrule
         
    \end{tabular}
    \label{tab:RQ3_res}
\vspace{-2.0em}
\end{table}

\section{Discussion\label{discussion}}
\subsection{Case Study}
In Section \ref{result}, we only study the human alignment of scores from various LLM-as-a-judge methods. However, output-based methods usually generate explanations for scores as well. Thus, we study two cases of scoring explanations here. Both judgments are generated by GPT-4o from group Vanilla in the output-based category.

\begin{figure}
    \centering
    \includegraphics[width=\textwidth]{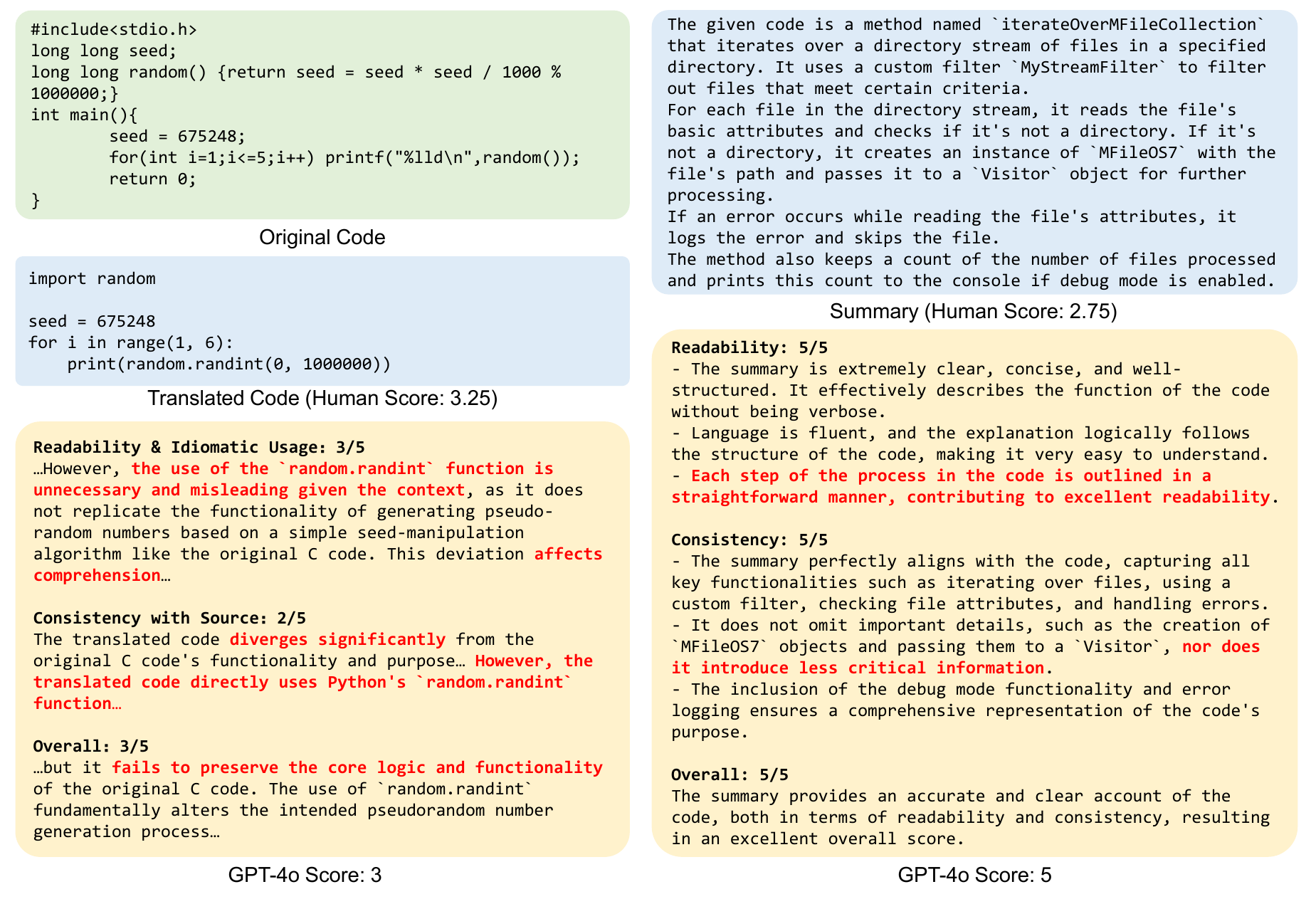}
    \vspace{-2.0em}
    \caption{Case study. The successful case from code translation is on the left while the failing case from code summarization is on the right.}
    \label{fig:case_study}
    \vspace{-1.0em}
\end{figure}

We present a passing case of code translation on the left side of Fig. \ref{fig:case_study}. In this example, the original C code implements a seeded custom pseudo-random function to generate random integers from 0 to 999999. However, the translated Python code simply invokes a similar function from the standard library without setting the seed, which clearly disregards the original purpose, and produces an incorrect range of \([0, 1000000]\). GPT-4o accurately identifies this discrepancy and highlights the failure to preserve the original intention, aligning with its moderate score of 3/5, although it overlooks the subtle difference in the range of possible random integers.

We also present a failing case of code summarization on the right side of Fig. \ref{fig:case_study}. Here, the responding LLM produces an overly detailed method summary, including excessive specifics such as identifier names and internal procedures, contrary to the instruction to summarize functionality without extensive details. Despite this, GPT-4o assigns a perfect score of 5/5, viewing the entire step-by-step explanation as essential information, while an ideal summary can be as concise as "The method iterates over a directory stream and filters out non-directory files." This case aligns with LLMs' verbosity bias introduced in related work \cite{DBLP:journals/corr/abs-2310-10076}.

\subsection{Implications of Findings}

\textbf{For developers:} Our findings indicate the potential of LLM-as-a-judge methods to replace human evaluators, to effectively evaluate the quality of LLM-generated content in certain SE tasks and save developers' time on selecting the best LLMs. We further conclude the following insights:

\begin{enumerate}

    \item Developers should carefully select LLM-as-a-judge methods, as their performance varies significantly across different categories of methods.

    \begin{itemize}
    
        \item Output-based methods with large LLMs like GPT-4o or DeepSeek-V2.5 offer the most human-aligning evaluation with proper prompt and inference strategies. 
        
        \item Individual scoring should be preferred over pairwise comparison with current methods.

    \end{itemize}
    
    \item For different tasks, developers should leverage LLM-as-a-judge methods in diverse ways to exploit their strengths, as their performance is highly task-dependent:
    
    \begin{itemize}
    
        \item For code translation and generation, state-of-the-art methods demonstrate mid-to-strong performance and can be used standalone, particularly when reference answers are unavailable or scoring explanations are required.

        \item For code summarization, LLM-as-a-judge methods should not be used directly or alone due to their insufficient alignment with human judgments. However, with carefully designed prompts to mitigate common LLM biases, they can still serve as a valuable complement to conventional metrics.
        
    \end{itemize}
    
\end{enumerate}



\textbf{For researchers:} Our study reveals the effectiveness and limitations of LLM-as-a-judge methods in SE tasks and shows some potential future directions, specifically:

\begin{enumerate}

    \item Current methods lack generalizability across SE tasks, as evidenced by their task-dependent performance:
    
    \begin{itemize}
    
        \item While SFT methods for evaluation exist, their performance on SE tasks is likely limited by the absence of challenging SE-specific data in training sets. Future research could benefit from curating difficult, SE-specific human preference datasets for fine-tuning smaller LLMs. Instructions in these datasets can originate from challenging benchmarks or even complicated real-world scenarios. It is also important to design strategies for state-of-the-art LLMs to generate responses and human-like judgments.

        \item Non-SFT methods use uniform prompt formats and inference strategies across tasks, without task-specific adaptations or utilizing code-specific features or structures. We therefore propose that designing SE- or even task-specific evaluation methods may yield more accurate and robust results than general-purpose evaluation frameworks.
        
    \end{itemize}

    \item There are still gaps to be bridged between LLM and human evaluators:

    \begin{itemize}
    
        \item During evaluation, LLMs typically rely solely on the predefined evaluation criteria. In contrast, human evaluators can compare multiple responses, implicitly identifying common strengths and weaknesses to streamline the evaluation process. To bridge this gap, researchers could enhance LLM-as-a-judge methods by asking LLMs to summarize insights from previous evaluation sessions. These insights could be included in the prompt to provide additional evaluation context. Multiple responses in the prompt are also valuable for LLMs to make comparisons.

        \item Human evaluators often discuss and reach a consensus, whereas LLM-as-a-judge frameworks typically contain a single LLM instance. To bridge this gap, researchers could develop multi-agent evaluation systems, where multiple LLM instances evaluate responses from different perspectives. This approach would enable more comprehensive and nuanced evaluations, akin to collaborative human judgment.

        \item The underperformance of LLM-as-a-judge in evaluating code summaries reveals a critical misalignment between benchmark task definitions and LLM interpretations. In our experiments, CodeXGLUE expects concise, docstring-style summaries while LLMs default to detailed explanations due to their verbosity bias. To improve evaluation reliability, researchers should mitigate LLMs' implicit, bias-influenced assumptions about the task to ensure they correctly understand task objectives.
        
    \end{itemize}
\end{enumerate}



\section{Conclusion\label{conclusion}}
In this paper, we empirically investigate the effectiveness of different types of LLM-as-a-judge methods on three SE datasets. We generate and manually score LLM responses, and assess these methods' alignment with human scores. Our results indicate that these methods demonstrate task-dependent performance, ranging from near-human to unusable when scoring individual responses, and generally perform worse in pairwise comparisons. We further analyze score characteristics, discovering that the most human-aligning methods display a balanced human-like distribution. Finally, we discuss key findings and implications for future development and application of LLM-as-a-judge in SE evaluation, hoping that these insights can assist future research in this area.

\section*{Data Availability}
Our source code and data is publicly available at \cite{replication-package}.

\bibliographystyle{ACM-Reference-Format}
\bibliography{references}

\end{document}